\documentclass[a4paper,fleqn,usenatbib]{mnras}

\usepackage[T1]{fontenc}
\usepackage{ae,aecompl}
\usepackage{graphicx}	% Including figure files
\usepackage{amsmath}	% Advanced maths commands
\usepackage{amssymb}	% Extra maths symbols
\title[A Supra-galactic Globular Clusters Conundrum]{A Supra-galactic Conundrum: The Globular Clusters Colour Pattern in Virgo Galaxies}

\author[Forte et al.]{Juan
  C. Forte$^{1,2}$\thanks{E-mail: planeta.jcf@gmail.com}, Sergio A. Cellone$^{2,3,6}$, Mar\'ia E. De Rossi$^{7,8}$, Carlos Escudero$^{2,3,4}$,  
  \newauthor 
Favio R. Faifer$^{2,3,4}$, Douglas Geisler$^{5,9,10}$, N\'elida M. Gonz\'alez$^{3}$, Mar\'ia C. Scalia$^{3}$, 
  \newauthor
Leandro Sesto$^{2,4}$, Anal\'ia V. Smith Castelli$^{2,4}$, E. Irene Vega$^{2,3}$\\
$^1${Instituto Argentino de Matem\'atica "Alberto P. Calder\'on", CONICET, Saavedra 15, 1083, CABA, Argentina}\\
$^2${Consejo Nacional de Investigaciones Cient\'ificas y T\'ecnicas, Godoy Cruz 2290, C1425FQB, CABA, Argentina}\\
$^3${Facultad de Ciencias Astron\'omicas y Geof\'isicas, Universidad
  Nacional de La Plata, Paseo del Bosque, B1900FWA, La Plata, Argentina}\\
$^4${Instituto de Astrof\'isica de La Plata (CCT-La Plata, CONICET-UNLP),
  Paseo del Bosque, B1900FWA, La Plata, Argentina}\\
$^5${Departamento de Astronom\'ia, Universidad de Concepci\'on, Casilla 160-C, Concepci\'on, Chile}\\
$^6${Complejo Astron\'omico "El Leoncito" (CASLEO), CONICET-UNLP-UNC-UNSJ, San Juan, Argentina}\\
$^7${Universidad de Buenos Aires, FCEN y Ciclo B\'asico Com\'un, Buenos Aires, Argentina}\\
$^8${CONICET-Universidad de Buenos Aires, Instituto de Astronom\'ia y F\'isica del Espacio, Buenos Aires,
Argentina}\\
$^9${Instituto de Investigaci\'on Multidisciplinario en Ciencia y Tecnolog\'ia, Univ. de La
Serena. Avenida Ra\'ul Bitr\'an S/N, La Serena, Chile}\\
$^{10}${Departamento de F\'isica y Astronom\'ia, Facultad de Ciencias, Univ. de La Serena. Av.
Juan Cisternas 1200, La Serena, Chile}
}
\date{Accepted October 8, 2018.}
\pubyear{2018}
\begin{document}
\label{firstpage}
\pagerange{\pageref{firstpage}--\pageref{lastpage}}
\maketitle
% Abstract of the paper
\begin{abstract}
The presence of systematic modulations in the colour distributions in composite samples of globular clusters associated with
 galaxies in the Virgo and Fornax clusters has been reported in a previous work. In this paper we focus on the 27 brightest galaxies in Virgo,
 and in particular on $NGC~4486$, the dominant system in terms of globular cluster population. The new analysis includes $\approx$ 7600 cluster
 candidates brighter than $g=$24.5 (or $T_{1}$$\approx$ 23.70). The results indicate the presence of the characteristic Virgo pattern in these galaxies and that 
 this pattern is detectable over a galactocentric range from 3 to 30 $Kpc$ in $NGC~4486$. This finding gives more support to the idea that the pattern has 
 been the result of an external, still not identified phenomenon, capable of synchronizing the cluster formation in a kind of viral process, and on supra-galactic scales 
 (also having, presumably, an impact on the overall star formation history in the entire Virgo cluster).
\end{abstract}
%%%%%%%%%%%%%%%%% BODY OF PAPER %%%%%%%%%%%%%%%%%%
\section{Introduction}
\label{Intro}
%****************************************************************************
%This is a simple template for authors to write new MNRAS papers.
%See \texttt{mnras\_sample.tex} for a more complex example, and \texttt{mnras\_guide.tex}
The so called globular clusters ($GCs$) colour "bi-modality" has been a dominant paradigm for
almost 25 years. In short, the term describes that in general, and with  a few exceptions \citep[see, for example,][]{Sesto2016}
the integrated colour distributions of $GCs$ in bright galaxies are 
dominated by two cluster populations: "blue" and "red" $GCs$ \citep[e.g.][]{Brodie2006, Harris2017}.

A first step towards exploring this issue beyond bi-modality was presented in \citet{Forte2017}
 ($F2017$ in what follows), who performed an analysis of composite $GC$ samples associated with
 galaxies in the Virgo and Fornax galaxy clusters. That work introduced an elementary pattern recognition
 technique, based on the {\bf frequency} of given colours in composite samples of $GCs$ in different galaxies.

This approach, using the $ACS$ photometry by \citet{Jordan2009} and \citet{Jordan2015}, led
 to the discovery of distinct modulations in the $(g-z)_{o}$ (reddening corrected) $GC$ colours in both galaxy clusters,
 that were termed as the $Template~Virgo~Pattern$ and $Template~Fornax~Pattern$ ($TVP$ and $TFP$,
 respectively). The Virgo pattern is characterized  by the $(g-z)_{o}$ colours listed in Table~\ref{table_1} of F2017 
 (namely, 0.74, 0.85, 0.95, 1.05, 1.13, 1.21, 1.29, 1.39, 1.48, 1.60 and 1.72). Values in brackets along
 the following text identify a given colour in the $TVP$.

The basic idea is that composite $GC$ samples may  enhance the presence of possible common features as,
for example, "peaks" or "valleys" in the  $GCs$ colour distribution ($GCCD$ in what follows). The colour 
 modulations are most evident for $GCs$ in moderately bright early type galaxies, with $M_{g}$ from -20.2 to
 -19.2, but they were also found in some $GC$ $\bf sub-samples$ of the brightest galaxies in these clusters  
(e.g. $NGC~4474$ and $NGC~4486$ in Virgo; $NGC~1399$ and $NGC~1404$ in Fornax).
 The $TVP$ is in fact a "die-hard" feature as it survives after splitting the $GCs$ sample in terms of cluster
brightness, galaxy brightness or spatial position as described in $F2017$.  

After rejecting the incidence of field contamination, instrumental, and statistical effects, the conclusion in $F2017$ was that
the colour patterns are not spurious but have a $\bf physical~entity$. On statistical grounds, Monte Carlo models
 along the lines described in \citet*{Forte2007} ($FFG07$ in what follows), indicate that the probability of
 having a colour pattern with the properties of the $TVP$ as the result of statistical fluctuations in a $GC$ sample with 
 $\approx$ 1500 objects, is practically null (i.e., below 0.001). 

This paper focusses on $GCs$ in the brightest Virgo cluster galaxies and extends the previous uni-dimensional
analysis (colour) to a bi-dimensional space (colour-magnitude). In particular, we study
the 27 brightest galaxies in Virgo ($M_{g}$ from -23.0 to -19.2 ) since, for fainter galaxies, the $GCs$ 
colour pattern is not evident and appears only marginally in some cases.

 The galaxy sample now includes 12 giant galaxies (brighter than $M_{g}=$-20.2), not discussed in $F2017$, and among them, $NGC~4486$,
 the dominant system in terms of the $GC$ population. The outer regions of this system are explored using previously published Washington
photometry ($FFG07$).
%------------------------------------------------------------------------------------------------
\section{Revisiting the Template Virgo Pattern.}
\label{sec2}

In this section we revisit the so called $Template~Virgo~Pattern$ using the same pattern recognition technique presented in
$F2017$ although introducing two changes in the strategy of the data analysis.

First, by performing an analysis of the $HST-ACS$  photometric errors given in \citet{Jordan2009}, we found that the colour pattern can be more clearly detected adopting a limiting magnitude $g_{o}=$ 24.5, i.e., half a magnitude brighter than the cut-off in $F2017$. With this magnitude limit, the "colour spread function" ($CSF$; the function that results after combining photometric errors and the smoothing gaussian kernel), allows the resolution of colour peaks at a level of $\approx$ 0.07 mags. in $(g-z)_{o}$.

Second, the finding routine was run on each of the 27 brightest galaxies in Virgo, instead of grouping galaxies within a sampling
window defined in the colour vs. absolute magnitude diagram. This approach eventually removes the chance that the observed pattern is a result of the type of adopted $GC$ sampling.

The galaxy sample is listed in Table~\ref{table_1} and is ordered by decreasing brightness, as in Table 1 of \citet{Ferrarese2006},
 who present structural photometric parameters for these galaxies. Table~\ref{table_1} gives the galaxy identification, $(g-z)_{o}$ colour shifts (see below), galactocentric range, and number of sampled $GCs$ with  $g_{o}$=20.0 to 24.5.

We remark that the $GC$ sample in Table~\ref{table_1} is a factor of three times larger than that defined by clusters in the moderately bright galaxies ($M_{g}$ from -20.2 to -19.2) discussed in $F2017$. 

Running the colour peak finding routine ($PFR$ in what follows) on each individual galaxy shows that, as a general trend in giant galaxies, the colour patterns are difficult to find within galactocentric radii ($R_{gal}$) smaller than 40$\arcsec$. For these galaxies we set a galactocentric search range from 40\arcsec to 100\arcsec. For the remaining ones, the galactocentric domain was set from 0\arcsec to 100\arcsec. 

\begin{figure}
\vspace{-0.5cm}
	\includegraphics[width=\columnwidth]{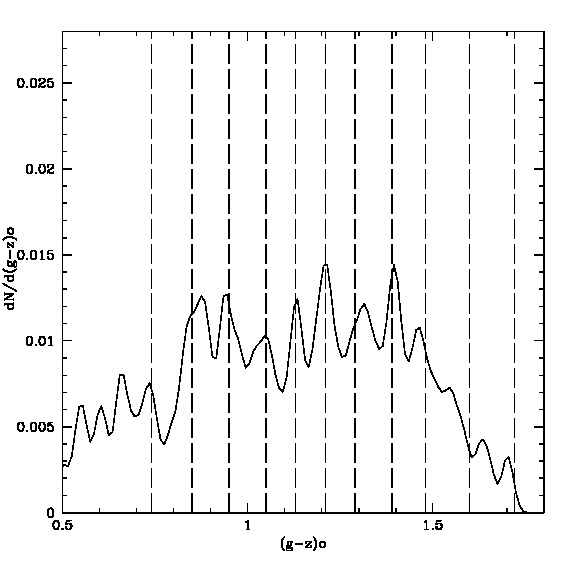}
    \caption{Smoothed relative $(g-z)o$ colour frequency corresponding to 311 peaks found on the $GC$ sample belonging
 to the 27 galaxies listed in Table~\ref{table_1}. The cluster colours are the original photometric data (corrected by interstellar extinction).
 Vertical lines indicate to the $TVP$ colours. 
}
    \label{fig:fig1}
\end{figure}

The results from the $PFR$ reveals the presence of 311 colour peaks (corresponding to a total of 4289 GCs) whose smoothed distribution, adopting a Gaussian kernel of 0.015 mag. (as in $F2017$), is shown in Fig.~\ref{fig:fig1}. This diagram, normalized by the total number of colour peaks, indicates the relative frequency of each colour peak (irrespective of the number of $GCs$  with that colour) and shows nine coincidences with the $TVP$. The differences between each colour peak and that of the nearest colour in the $TVP$, leads to a $rms=$0.016.
%-----------------------------------------------------------------------------------------------------
%----------------------------------------------------------------------------------------------------
The situation is poorly defined regarding the reddest $TVP$ features, at [1.60] and [1.72].

The nature of the peaks outside the $GCs$ colour range is difficult to asses. In an speculative way, these  features could arise in field objects as they do not show any obvious spatial concentration towards the galaxy centres.

The outputs from the $PFR$ also indicate that, in some cases, the whole $TVP$ is detectable although with some small shifts in colour.
$NGC~4472$, the brightest Virgo galaxy, is a good example as its $GCCD$ shows all the $TVP$ features, although shifted by +0.05 mag in
$(g-z)o$. Similar systematic shifts are also detectable in other galaxies with well populated $GCs$ systems.

 The shifts that bring a given pattern into agreement with the $TVP$ (by minimizing the $rms$ of the differences between each colour peak and the nearest one in the $TVP$), are listed in Table~\ref{table_1}.
 
As noted in $F2017$, these shifts are within  the uncertainties in the adopted interstellar reddening. In fact, the $rms$ of the original infra-red emission $vs.$ colour excess map calibration  by \citet*{Schlafly2011} is close to $\pm$ 0.03 mag. in $E(B-V)$ that 
corresponds to $\pm$ 0.05 mag. in $(g-z)$.

\begin{figure}
\vspace{-0.5cm}
	\includegraphics[width=\columnwidth]{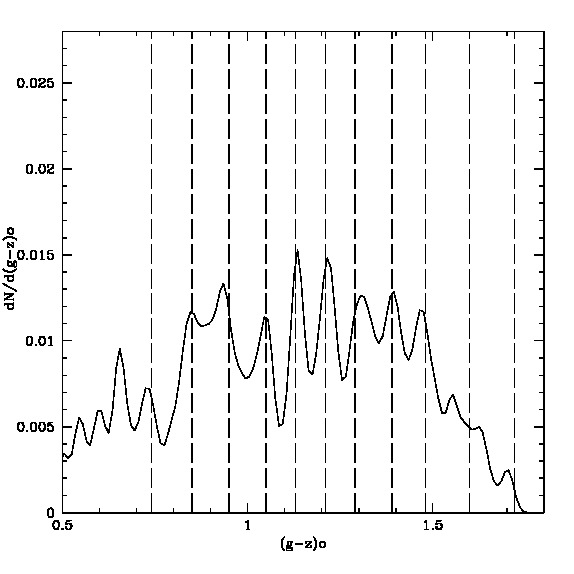}
    \caption{Smoothed relative $(g-z)o$ colour frequency corresponding to 311 peaks found on the $GC$ sample belonging
 to the 27 galaxies listed in Table~\ref{table_1}. Some of the cluster colours (corrected by interstellar extinction) have been shifted 
 according to the values listed in that table. Vertical lines indicate the $TVP$ colours.
}
    \label{fig:fig2}
\end{figure}

The results from the $PFR$, after adopting the colour shifts listed in Table~\ref{table_1}, lead to the relative frequency distribution displayed in 
Fig.~\ref{fig:fig2}. This diagram, compared to the previous one, shows an overall increase of the contrast of the peaks and also a decrease of the $rms$ value to 0.010.

%-----------------------------------------------------------------------------------------------------
%----------------------------------------------------------------------------------------------------

\section{The composite Globular Clusters Colour Distributions in the 27 brightest Virgo galaxies.}
\label{section3}
In what follows we present the $GCCDs$ (normalized by total numbers of $GCs$), both in the discrete and smoothed histogram versions. The colour-bin size (0.04 mag) and the Gaussian kernel (0.015 mag) are those discussed in $F2017$. 

The differences between the colour peaks found by the pattern recognition routine and those of the nearest colour in the $TVP$ definition, leads to a $rms$ value that is an indicator of the agreement between  
both patterns. This $rms$ is calculated for colour peaks within the colour range $(g-z)_{o}$ from 0.80 to 1.60, characteristic of old $GCs$.

In order to derive the  composite $GCCDs$, the clusters belonging to the galaxies listed in Table~\ref{table_1}  were  combined  in  three different groups (separated by blank lines in the table) containing approximately the same number of $GCs$ each.

\begin{figure}
\vspace{-0.5cm}
	\includegraphics[width=\columnwidth]{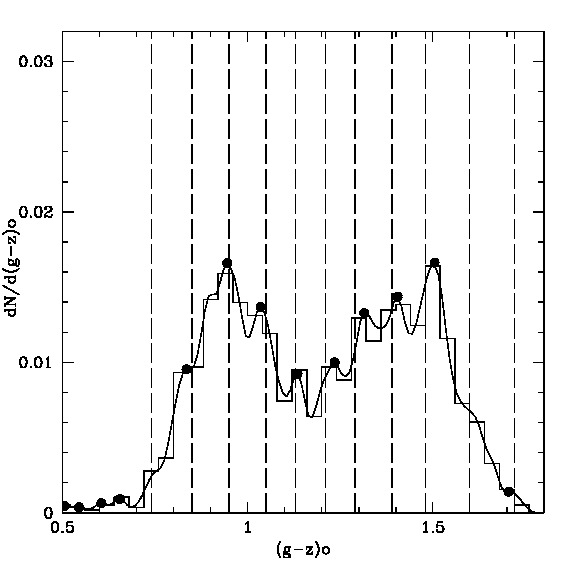}
    \caption{$(g-z)o$ colour distribution corresponding to a composite $GC$ sample that includes  1447 objects in the three
 brightest Virgo giants: $NGC$ 4472, 4486 and 4649 and within a galactocentric range from 40$\arcsec$ to 100$\arcsec$. The $GC$ colours have been shifted according to Table~\ref{table_1}. Vertical lines correspond to the $TVP$ colours. Dots indicate the colour peaks found by the $PFR$.
}
    \label{fig:fig3}
\end{figure}

The $GCCD$ corresponding to galaxies brighter than $M_{g}=$-21.8, including  $NGC~4472,~4486$ and~$4649$ (with 1447 $GCs$), is displayed in Fig.~\ref{fig:fig3}. The vertical lines in this figure (and in following ones) correspond to the $TVP$ colours, while dots indicate the colour peaks found by the $PFR$. The $rms$ corresponding to the differences between the detected peaks and those that define the $TVP$ is $\pm$0.016 (eight peaks).
%-----------------------------------------------------------------------------------------
%------------------------------------------------------------------------------------------

In this case the  feature at [0.85] is barely detectable. This is consistent with results available in the literature  \citep[see, for example,][]{Escudero2018} showing
that the inner regions of giant galaxies are usually dominated by red $GCs$ while the bluer globulars are less abundant.

\begin{figure}
\vspace{-0.5cm}
	\includegraphics[width=\columnwidth]{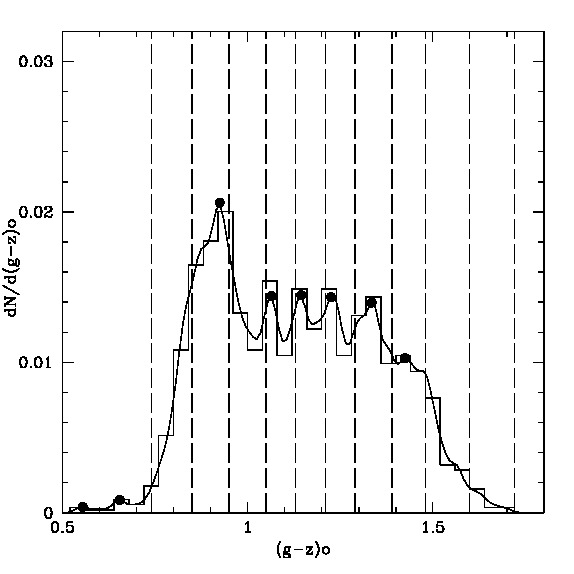}
    \caption{$(g-z)o$ colour distribution corresponding to a composite sample of 1411 $GCs$ belonging
 to nine giant galaxies in Table~\ref{table_1} (from $NGC$ 4406 to 4762). Vertical lines correspond to the $TVP$ colours. Dots indicate the colour peaks found by the $PFR$.
}
    \label{fig:fig4}
\end{figure}

Fig.~\ref{fig:fig4} is the $GCCD$ corresponding  to nine less massive giants ($M_{g}=$-20.2 to -21.8; from $NGC$ 4406 to 4762 in Table~\ref{table_1};  1411 GCs). This group has a slightly larger rms  ($\pm$0.018 mag) compared with the previous one ($\pm$ 0.016), and again, the colour peak at [0.85] is not evident.

As expected, the reddest $TVP$ peaks in this galaxy group are less prominent than those in the more massive, and more chemically enriched, ones displayed in  Fig.~\ref{fig:fig3}.
%----------------------------------------------------------------------------------------
%--------------------------------------------------------------------------------------
 
\begin{figure}
\vspace{-0.5cm}
	\includegraphics[width=\columnwidth]{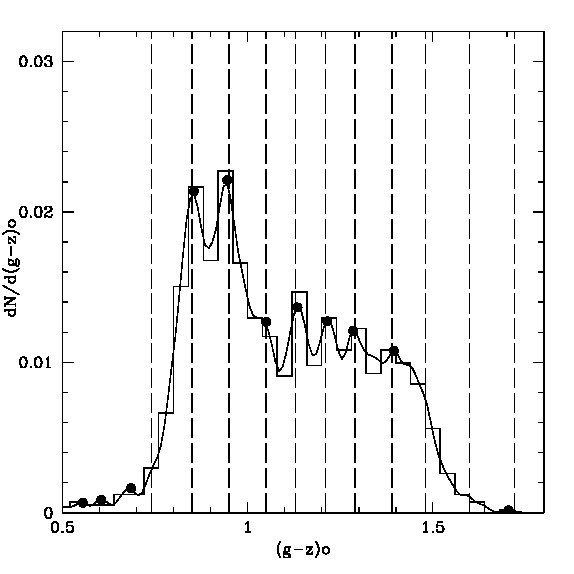}
    \caption{$(g-z)o$ colour distribution for 1431 $GCs$ corresponding to fifteen moderately bright galaxies listed
 in Table~\ref{table_1} (from $NGC$ 4459 to 4578). Vertical lines correspond to the $TVP$ colours. Dots indicate the colour peaks found by the $PFR$.
}
    \label{fig:fig5}
\end{figure}

Finally, the $GCCD$ corresponding to 1431 $GCs$ in fifteen moderately bright galaxies ($M_{g}=$-19.2 to -20.2) is depicted in Fig.~\ref{fig:fig5}. This figure shows all the $TVP$ peaks  from [0.85] to [1.39] and yields an $rms$ of only 0.007 mag. This group shows a prominent blue peak at [0.85]. This can be expected since, being smaller in angular size, their outer regions fall within the real coverage of the $ACS$ photometry.

The determination of the $TVP$ colours in $F2017$ was based on a galaxy sample that included galaxies fainter than Mg=-20.2. However, this section shows that this pattern can be also consistently recognized in the composite $GCCD$ corresponding to the twelve giant galaxies listed in Table~\ref{table_1}. Furthermore, the two brightest galaxies in Virgo ($NGC 4472$, and$~4486$) have  individual $GCCDs$ where the $TVP$ pattern is easily recognizable (see below).

The presence or absence of some colour peaks, at the blue or red edges of the $GCCDs$, can be explained in terms of the different chemical enrichment levels that characterize each galaxy group. In general,
the most massive galaxies (with high chemical enrichment) will show red $TVP$ components that become less evident or absent in lower mass galaxies.

\begin{figure}
\vspace{-0.5cm}
	\includegraphics[width=\columnwidth]{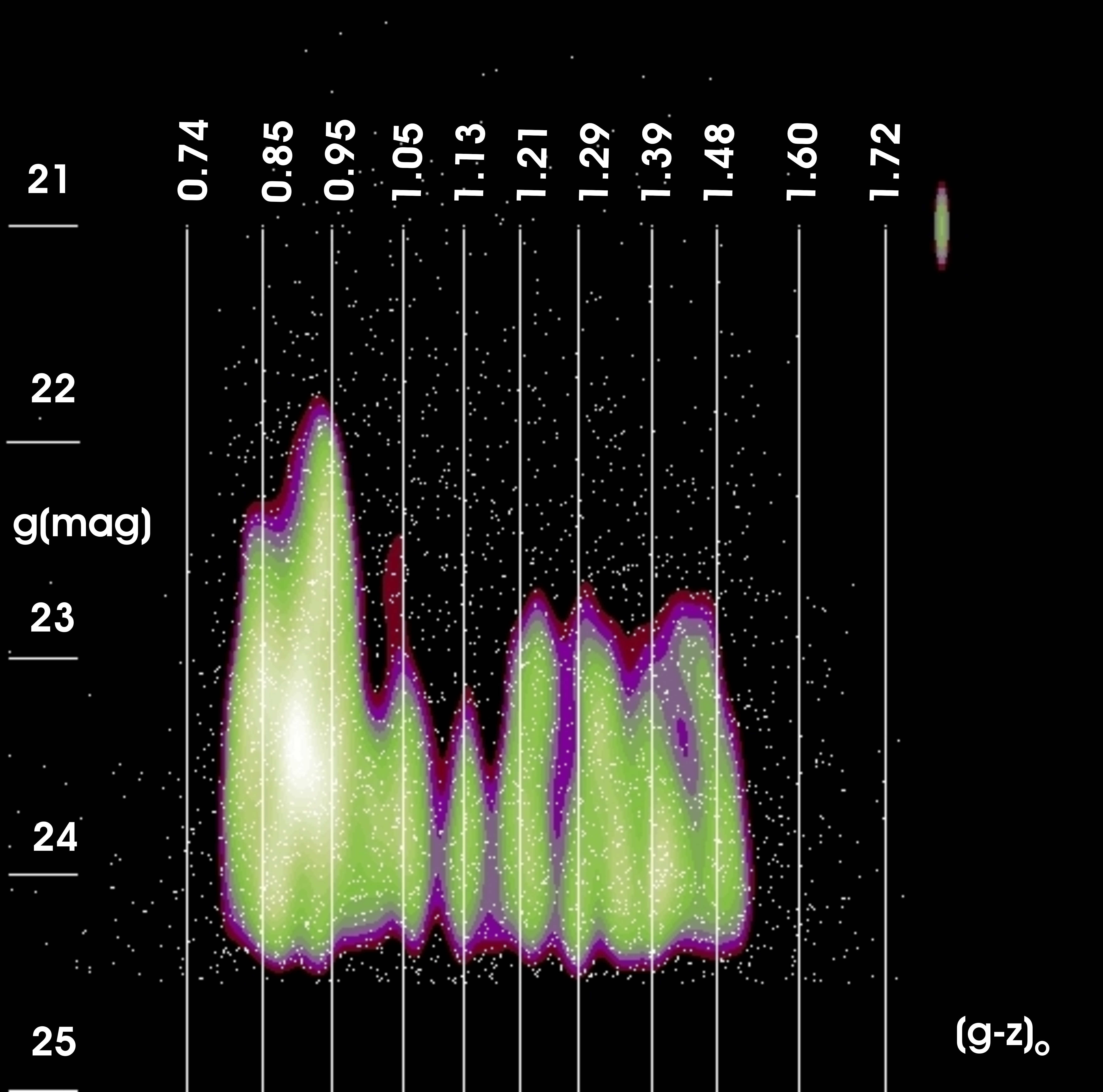}
    \caption{Smoothed colour-magnitude diagram for 4289 $GCs$ corresponding to all the galaxies listed
 in Table 1 (within the indicated galactocentric ranges, and adopting the indicated colour shifts). 
 The vertical lines correspond to the $TVP$ colours. White dots represent individual $GCs$. The contour of the
 convolving kernel is shown at upper right.
}
    \label{fig:fig6}
\end{figure}

 Conversely, low mass galaxies will exhibit blue $TVP$ components (that are less prominent in the inner regions of massive galaxies). 
%----------------------------------------------------------------------------------
%--------------------------------------------------------------------------------------
 The smoothed colour magnitude diagram for the 4289 $GCs$ sample corresponding to all the galaxies listed in Table~\ref{table_1}, is shown in Fig.~\ref{fig:fig6}.

For constructing this diagram, the $g_{o}$ magnitudes and $(g-z)_{o}$ colours were mapped as components of a $501\times501$  matrix defined between $g_{o}=$~20.0 and 25.0, and $(g-z)_{o}$ from 0.5 to 2.0. The $IRAF$ routine $gauss$ was run on this matrix, generating a $FITS$ image after adopting a colour kernel of 0.015 mag, and a $g$ magnitude kernel of 0.25 mag. The first value is that used in the $F2017$ uni-dimensional analysis while the second one was set after $trial-and-error$ runs (with $g_{o}$ magnitude kernels ranging from 0.10 to 0.50 mag) aiming at enhancing the presence of the colour pattern.

Fig.~\ref{fig:fig6} is a false-colour display of the $FITS$ image (after adopting a proper minimum/maximum range and contrast) and shows a highly structured $GCCD$. The contour of the Gaussian kernel is shown at the upper right in this diagram.
%---------------------------------------------------------------------------------------
%----------------------------------------------------------------------------------------
The expectation is that, if the colour patterns are not random effects, they should show some coherence in the colour magnitude plane. In fact, Fig.~\ref{fig:fig6} indicates that the colour modulations are not the result of localized data "clumps" but, rather, distinct structures spanning different ranges in apparent magnitude and coincident with the $TVP$ colours, shown as vertical lines (see Section 6). The "widths" of these structures are in fact set by a combination of the $CSF$ and the convolving kernel.

A remarkable feature in the last diagram is the presence of two blue peaks [0.85; 0.95] that appear in most of the individual galaxies listed in  Table~\ref{table_1}.
 
%------------------------------------------------------------------------------
\begin{table}
\centering
\caption{$GCs$ samples ($g_{o}$=20.0 to 24.5) in the 27 brightest Virgo galaxies.}
\begin{tabular}{c c c c c c c}
\hline
\hline
\textbf{NGC}& &\textbf{$(g-z)_{o}$~shift}& &\textbf{Gal.~range $\arcsec$}& &\textbf{N of GCs}\\
\hline
 4472& &-0.05& &40-100& &330\\
 4486& &-0.00& &40-100& &762\\
 4649& &-0.02& &40-100& &355\\
\\
 4406& &+0.05& &40-100& &172\\
 4382& &+0.03& &40-100& &176\\
 4374& &-0.02& &40-100& &203\\
 4365& &~0.00& &40-100& &304\\
 4526& &+0.02& &40-100& &~97\\
 4621& &+0.01& &40-100& &144\\
 4552& &-0.02& &40-100& &176\\
 4473& &~0.00& &40-100& &~99\\
 4762& &~0.00& &40-100& &~40\\
\\
 4459& &~0.00& &0-100& &128\\
 4442& &~0.00& &0-100& &128\\
 4754& &~0.00& &0-100& &~61\\
 4267& &~0.00& &0-100& &129\\
 4371& &~0.00& &0-100& &120\\
 4570& &~0.00& &0-100& &~90\\
 4435& &~0.00& &0-100& &120\\
 4660& &~0.00& &0-100& &143\\
 4530& &~0.00& &0-100& &113\\
 4564& &~0.00& &0-100& &113\\
 4340& &~0.00& &0-100& &~29\\
 4417& &~0.00& &0-100& &~61\\
 4638& &~0.00& &0-100& &~58\\
 4478& &~0.00& &0-100& &~90\\
 4578& &~0.00& &0-100& &~48\\
\hline
\label{table_1}
\end{tabular}
\end{table}
%********************************************************************************************************
\section{The Template Virgo Pattern in NGC 4486}
\label{section4}
%--------------------------------------------------------------------------------
$NGC~4486$ is a well known paradigm of the so called "high $S_{n}$" galaxies \citep[e.g.][]{Tamura2006} and is the dominant system in Virgo in terms of its $GCs$ population. $F2017$  noted that the $TVP$ can be detected at some galactocentric radii and position angles ranges (see fig. 22 and 23 in that paper).

\begin{figure}
	\includegraphics[width=\columnwidth]{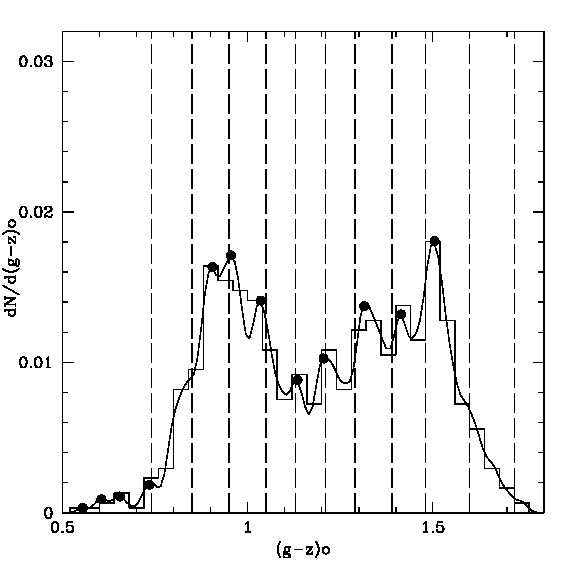}
\vspace{-0.5cm}
    \caption{Discrete bin and smoothed $(g-z)o$ colour distribution for 762 $GCs$ with $ACS$ photometry and
 galactocentric radii from 40\arcsec to 90\arcsec in $NGC~4486$. Vertical lines correspond to the $TVP$ colours.
 Black dots indicate the colour peaks found by the $PFR$.
}
    \label{fig:fig7}
\end{figure}

The presence of the $TVP$ is seen in Fig.~\ref{fig:fig7} where the $GCCD$ corresponds to 762 clusters with $ACS$ photometry and galactocentric radii from 40\arcsec to 90\arcsec. This diagram shows seven coincidences with the $TVP$ with an $rms$ of 0.019. A disagreement is seen for the bluest $GCs$ for which the [0.85] feature appears some 0.04 mag. redder than expected. The behaviour of the bluest $GC$ colours is similar to that described  in the previous discussion regarding blue $GCs$ in the inner regions of massive galaxies.
%---------------------------------------------------------------------------------------
%------------------------------------------------------------------------------------------ 

In  what follows we discuss the eventual detectability of the $TVP$ in the outer regions of $NGC~4486$ using the photometry presented by $FFG07$. These ground based observations were obtained with the the $Mayall$  $4-m$  telescope at $KPNO$, and adopting the $(C-T_{1})$ colour index defined  with these two passbands of the Washington photometric system.

A key issue in this context is a reliable  determination of the $(g-z)o$ vs $(C-T_{1})_{o}$ colour-colour relation, aiming at obtaining the equivalent $TVP$ colours in the $(C-T_{1})_{o}$ colour scale. This can be carried out using the $Gemini-griz$ photometry presented in \citet{Forte2013} for an offset field in $NGC~4486$. This  field includes 522 $GC$ candidates that also have Washington $C$ and $T_{1}$ photometry given in $FFG07$.
 
\begin{figure}
	\includegraphics[width=\columnwidth]{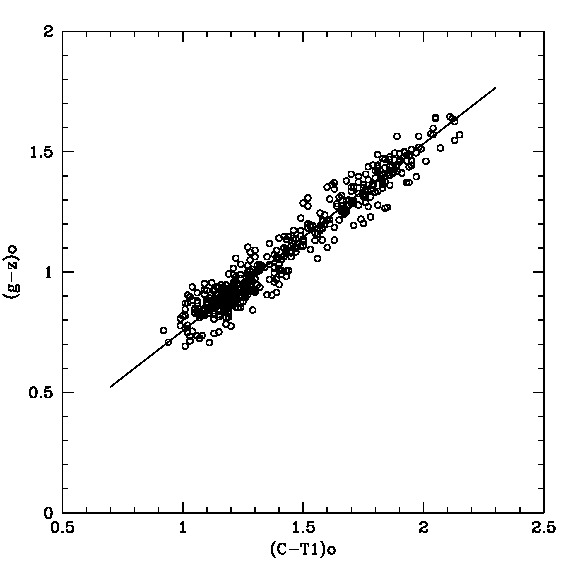}
\vspace{-0.5cm}
    \caption{The $(g-z)o$ vs. $(C-T_{1})_{o}$ colour relation. Dots correspond to 457 $GC$ candidates, taken
 from Forte et al. (2013), whose colours are compatible with those of old $GCs$ (see text).
}
    \label{fig:fig8}
\end{figure}

As a first step, an analysis of the $griz$ colours was performed in an attempt to remove field interlopers. This procedure left 457 objects with colours that are fully compatible with those of old $GCs$. The  $(g-z)o$ vs $(C-T_{1})_{o}$ colour-colour diagram for these objects is shown in Fig.~\ref{fig:fig8}. The straight line in this figure corresponds to a bi-sector fit (i.e., taking into account the photometric errors on both axes):
\begin{equation}
 (C-T_{1})_{o} = 1.287.(g-z)_{o} + 0.027
\end{equation}
(with coefficient uncertainties of $\pm$ 0.012 and $\pm$ 0.014, respectively).

The analysis of the residuals of this fit does not show any systematic trend or deviation from linearity with colour.

This relation then allows a link between the $(g-z)_{o}$ $TVP$ colours and those in the $(C-T_{1})_{o}$ colour scale, with uncertainties below $\pm$0.015 mag., yielding: 0.98, 1.12, 1.25, 1.38, 1.48, 1.58, 1.69, 1.82, 1.93, 2.09, 2.23.
%-----------------------------------------------------------------------------------------
%------------------------------------------------------------------------
%bf FROM HERE ON
\section{The GCCD in the outer regions of NGC 4486}
\label{section5}
\subsection{The GCCD in NGC 4486 for galactocentric radii between 90 and 240 arcsec.}
\label{ssection5a}
\begin{figure}
	\includegraphics[width=\columnwidth]{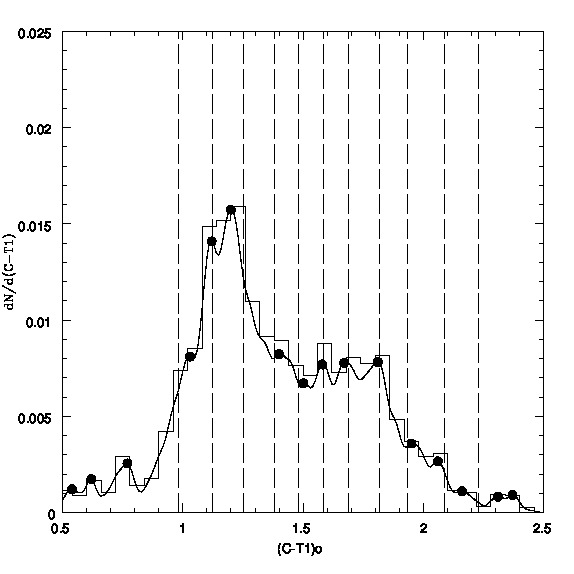}
\vspace{-0.5cm}
    \caption{Globular cluster $(C-T_{1})o$ colour distribution for 2048 objects with galactocentric radii from 90\arcsec to 240\arcsec and $T_{1o}=$21.0 to 24.0 in $NGC~4486$. Dots indicate the colour peaks detected by the $PFR$ routine. Vertical dashed lines indicate the $(C-T_{1})$ components of the $TVP$.
}
    \label{fig:fig9}
\end{figure}

 The $(C-T_{1})_{o}$ $GCCD$ for 2048 cluster candidates with galactocentric radii from 90$\arcsec$ to 240$\arcsec$ (with a complete areal coverage) 
 is displayed in Fig.~\ref{fig:fig9}. This sample corresponds to a magnitude range $T_{1o}$ from 20.0 to 24.0, i.e., some
 0.5 mag. brighter than the photometric cut-off in $FFG07$. The $PFR$ detects seven peaks compatible with the $TVP$ [in the colour range from $(C-T_{1})_{o}=$ 1.12 to 1.82] 
 with an $rms$ of 0.023 mag. (or 0.018 mag. in a $(g-z)$ colour scale). This value is comparable with those that characterize the $GCCDs$ in Virgo galaxies, as discussed 
 in Section 2 and 3. In fact, there is a clear similitude between Fig.~\ref{fig:fig9} and Fig.~\ref{fig:fig5}, corresponding to $GCs$ associated with the moderately bright galaxies.
 This similitude indicates that, as a whole, the $NGC 4486$ $GCs$ in this galactocentric range, have chemical abundances comparable to those in the bright galaxies sample listed in Table~\ref{table_1}.

 Fig.~\ref{fig:fig9} also shows a feature at $(C-T_{1})_{o}=$ 1.04 that has not been found in the other galaxies and whose nature is discussed in Section 6.
 
 Colour peaks redder than $(C-T_{1})_{o}=$ 1.82 are barely detectable and have not been included in the estimate of the $rms$ values (constrained to the $(C-T_{1})$ range from 0.93 to 1.87 in what follows). 
 Only 8 percent of the cluster population in this sample are redder than $(C-T_{1})=$1.87 (making difficult the eventual detection of the reddest peaks).
%--------------------------------------------------------------------------
%---------------------------------------------------------------------------

The $GCs$ sample within this region was analysed as a function of the $T_{1o}$ magnitude aiming at detecting the 
 $GCs$ magnitude range where, eventually, the colour pattern would be better defined.

\begin{figure}
	\includegraphics[width=\columnwidth]{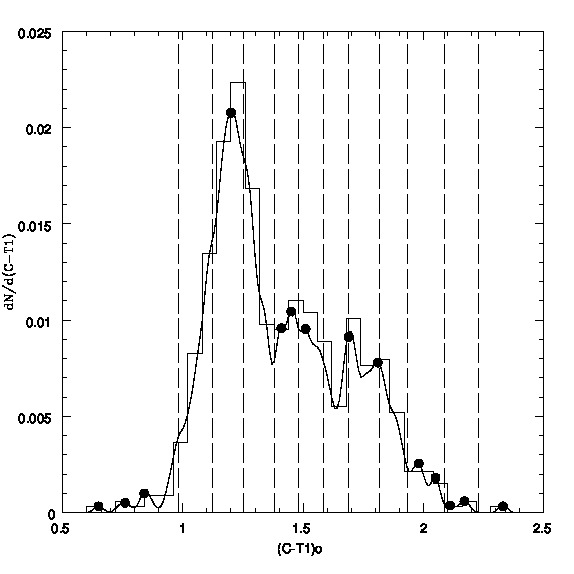}
\vspace{-0.5cm}
    \caption{Globular cluster $(C-T_{1})o$ colour distribution for 588 objects with galactocentric radii from 90\arcsec to 240\arcsec and $T_{1o}=$20.0 to 22.0, in $NGC~4486$. 
Dots indicate the colour peaks detected by the $PFR$ routine. Vertical dashed lines indicate the $(C-T_{1})$ components of the $TVP$. 
}
    \label{fig:fig10}
\end{figure}

 Fig.~\ref{fig:fig10} is the $GCCD$ corresponding to 588 clusters with $T_{1o}=$20.0 to $T_{1o}=$22.0 These objects probably include a fraction of 
$UCD$ objects overlapped with the brightest $GCs$ \citep[e.g.][]{Zhang2018}. This sample is clearly dominated by blue $GCs$ and the pattern features are poorly defined ($rms=$ 0.034).
%--------------------------------------------------------------------------
%---------------------------------------------------------------------------
 
\begin{figure}
	\includegraphics[width=\columnwidth]{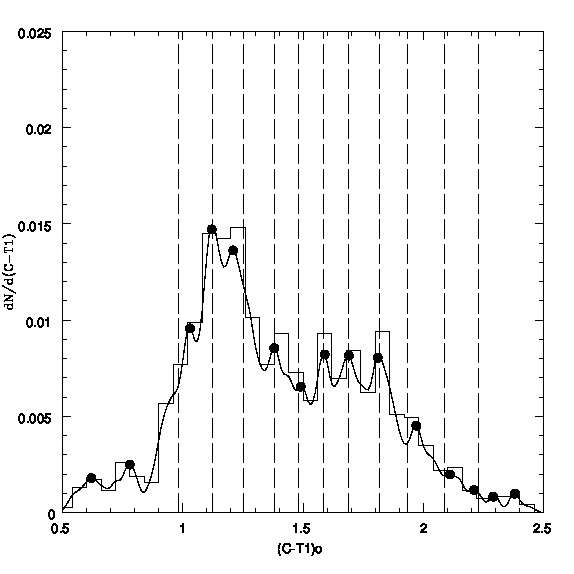}
\vspace{-0.5cm}
    \caption{Globular cluster $(C-T_{1})o$ colour distribution for 1240 objects with galactocentric radii from 90\arcsec to 240\arcsec and $T_{1o}=$22.0 to 23.7, in $NGC~4486$. 
Dots indicate the colour peaks detected by the $PFR$ routine. Vertical dashed lines indicate the $(C-T_{1})$ components of the $TVP$.
}
    \label{fig:fig11}
\end{figure}

After adopting different ranges in the $T_{1o}$ magnitudes, the best definition of the $TVP$ corresponds to clusters with $T_{1o}=$22.0 to $T_{1o}=$23.7 ($rms=$ 0.019), shown in Fig.~\ref{fig:fig11}, and includes 1240 $GCs$. The [1.04] feature is clearly seen in this diagram.
%---------------------------------------------------------------------------------
%--------------------------------------------------------------------------------

\begin{figure}
	\includegraphics[width=\columnwidth]{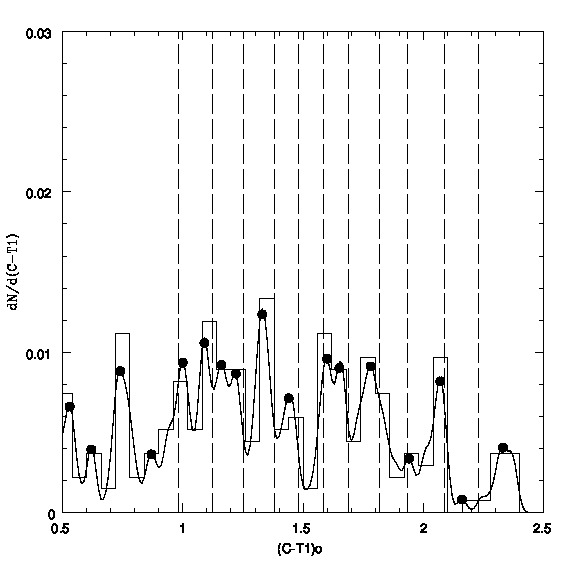}
\vspace{-0.5cm}
    \caption{Globular cluster $(C-T_{1})o$ colour distribution for 242 objects with galactocentric radii from 90\arcsec to 240\arcsec and $T_{1o}=$23.7 to 24.0, in $NGC~4486$. Dots indicate the colour peaks detected by the $PFR$ routine. Vertical dashed lines indicate the $(C-T_{1})$ components of the $TVP$.
}
    \label{fig:fig12}
\end{figure}

 Finally, the faintest sample, corresponding to $T_{1o}=$23.7 to $T_{1o}=$24.00, and including 242 $GCs$, exhibits the $GCCD$ depicted in Fig.~\ref{fig:fig12} ($rms=$0.039). As expected, the $TVP$ does not seem to be detectable in this sample as a consequence of the larger photometric errors as well as for the presence of an increasing number of field interlopers.
%---------------------------------------------------------------------------
%----------------------------------------------------------------------------------------
\subsection{The GCCD in NGC 4486 for galactocentric radii between 240\arcsec and 390\arcsec.}
\label{ssection5b}
\begin{figure}
	\includegraphics[width=\columnwidth]{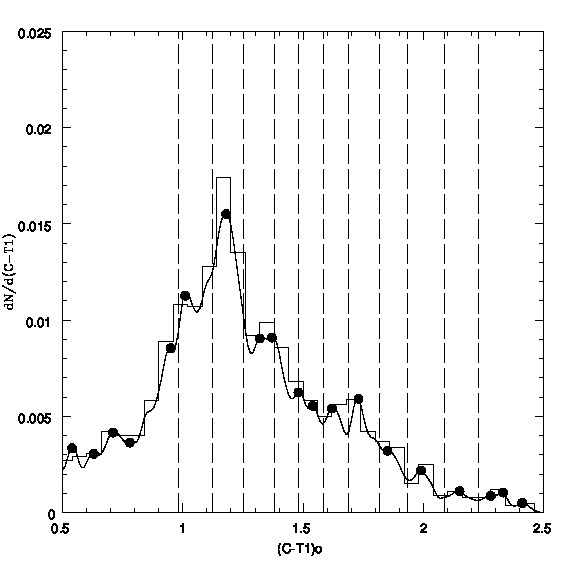}
\vspace{-0.5cm}
    \caption{$(C-T_{1})o$ $GGCD$ for 1800 clusters with galactocentric radii from 240\arcsec to 390\arcsec and
 $T_{1o}=$20.0 to 24.0 in $NGC~4486$. Dots are colour peaks found by the $PFR$. Vertical lines indicate the
$TVP$ colours.
}
    \label{fig:fig13}
\end{figure}

This region has an areal coverage of 80 percent in $FFG07$ and its $GCCD$, displayed in Fig.~\ref{fig:fig13}, shows a single dominat blue peak and a rather disordered pattern with a large $rms$ value of 0.035 mag. In principle, this result would indicate that the colour pattern is absent within that range of galactocentric radius.
%-----------------------------------------------------------------------------------------------------
%------------------------------------------------------------------------------------------------------

 A further test about this result was performed by analysing the $GCCDs$ on galactocentric annuli 30\arcsec wide. On these annuli, the $PFR$ generates the catalogue that allows the
 determination of the relative frequency of the colour peaks (a treatment similar to that described in Section 3), as well as the colour shift that brings the colour pattern within
  a given galactocentric annulus into agreement with the $TVP$ (by minimizing the colour $rms$ values as explained before). 

The number of $GCs$ within each annulus and the colour differences, are listed in Table~\ref{table_2}.
%------------------------------------------------------------------------------
\begin{table}
\centering
\caption{$GCs$ Number of globular clusters and colour differences between the $TVP$ and the patterns found in different
 annular rings in NGC 4486.}
\begin{tabular}{c c c c c c }
\hline
\hline
\textbf{Gal. range (arcsec)}& &\textbf{$(g-z)_{o}$~colour diff.}& &\textbf{N of GCs}\\
\hline
 90-120& &+0.02& &176\\
120-150& &+0.02& &149\\
150-180& &~0.00& &220\\
180-210& &~0.00& &180\\
210-240& &~0.00& &180\\
240-270& &+0.04& &179\\
270-300& &+0.08& &190\\
300-330& &+0.08& &190\\
330-360& &+0.09& &194\\
360-390& &+0.10& &195\\
\hline
\label{table_2}
\end{tabular}
\end{table}
%--------------------------------------------------------------------------------------------------------
 We note that the $FFG07$ photometry of $GCs$ in $NGC~4486$ has a complete areal coverage in $R_{gal}$ between 90$\arcsec$ and 390$\arcsec$ and position angles (N to E) from 90$\degr$ to 270$\degr$ (that corresponds to fifty percent of the total area) and that, in order to preserve an even coverage of the relative area of each annulus, the $PFR$ was run within these galactocentric limits. The $GCs$ magnitude range was set from $T_{1o}=$19.2 to 23.7 (corresponding to $g_{o}$$\approx$20.0 to 24.5).

\begin{figure}
	\includegraphics[width=\columnwidth]{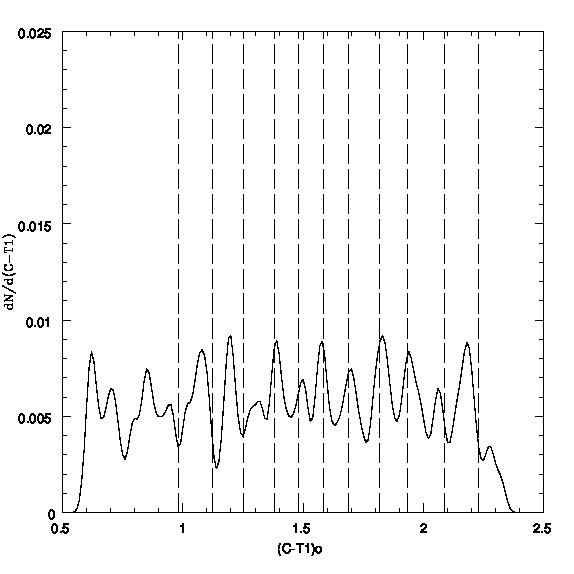}
\vspace{-0.5cm}
    \caption{Relative frequency of $(C-T_{1})_{o}$ colour peaks (without any colour shift) derived from data listed in Table 2. Vertical lines correspond to the $TVP$ colours. A disagreement between the $TVP$ colours is evident for peaks bluer than $(C-T_{1})_{o}$= 1.38.
}
    \label{fig:fig14}
\end{figure}

The relative frequency of the colour peaks  based on 1853 $GCs$, without any colour shift, is displayed in Fig.~\ref{fig:fig14}. Eight peaks redder than $(C-T_{1})_{o}=$1.38 have good coincidences with the $TVP$ colours ($rms=$$\pm$0.015). However, bluer peaks are in clear disagreement.
%--------------------------------------------------------------------------------------------------------

\begin{figure}
	\includegraphics[width=\columnwidth]{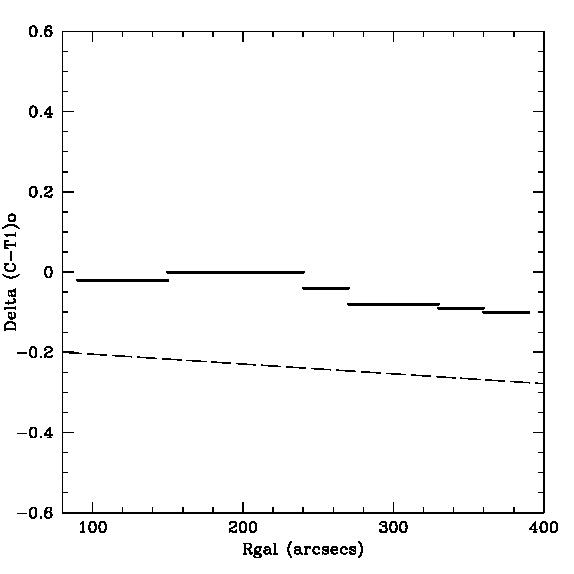}
\vspace{-0.5cm}
    \caption{$(C-T_{1})_{o}$ colour difference between the colour pattern observed within a
 given galactocentric range, indicated by the horizontal lines, and the $TVP$ colours. The dashed line (arbitrarily
 shifted downwards) corresponds to the $GC$ colour gradient mentioned in the text. 
}
    \label{fig:fig15}
\end{figure}

Fig.~\ref{fig:fig15} exhibits the colour differences listed in  Table~\ref{table_2} (with signs changed) plotted as a function of galactocentric radius. This figure indicates that the $TVP$ is detected with a consistent blueward trend as the galactocentric radius increases beyond $\approx$ 240$\arcsec$. Thus,  $GCs$ in this galactocentric range  seem to have a lower chemical abundance level compared to 
 those in the inner region of the galaxy (see previous Section).
$\bf{Tentatively}$, this blueward trend can be identified with the $GCs$ galactocentric colour gradient found by \citet{Forte2012}, indicated as a dashed line. This gradient corresponds to : $[\Delta(C-T_{1})$/$\Delta(R_{gal}$)]=2.6$\times$10$^{-4}$ mag./$\arcsec$ (for 12 Gy old $GCs$).

%-------------------------------------------------------------------------------------------------------------------------- 
%-----------------------------------------------------------------------------------------
\begin{figure}
	\includegraphics[width=\columnwidth]{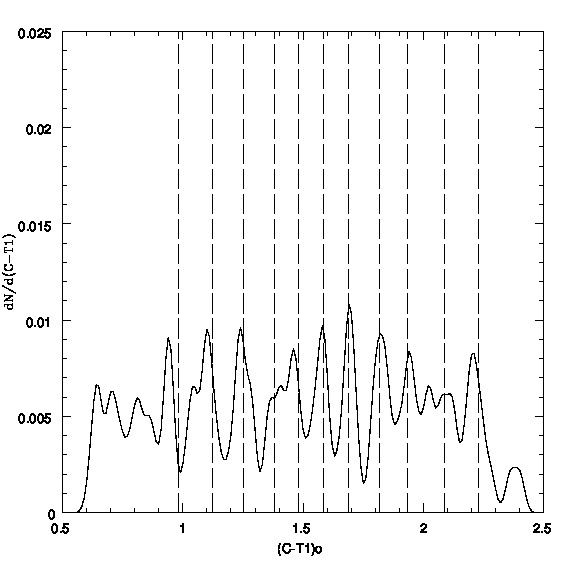}
\vspace{-0.5cm}
    \caption{Relative frequency of $(C-T_{1})_{o}$ colour peaks derived from data listed in Table 2. Vertical lines correspond
 to the $TVP$ colours. The $GC$ colours have been shifted according to the values indicated in the text.
}
    \label{fig:fig16}
\end{figure}
%-----------------------------------------------------------------------------------------
The adoption of the colour shifts listed in Table~\ref{table_2}, leads to the relative frequency of the colour peaks depicted in Fig.~\ref{fig:fig16}, where, on one side, the peaks increase their contrast (compared to those in Fig.~\ref{fig:fig14}) and, on the other, leads to a nearly complete correspondence with the $TVP$ colours(and a $rms$ of 0.018).\\

\begin{figure}
	\includegraphics[width=\columnwidth]{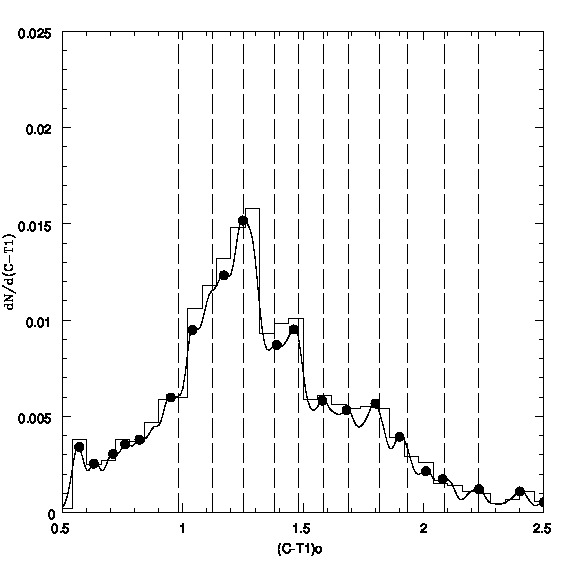}
\vspace{-0.5cm}
    \caption{Globular cluster $(C-T_{1})o$ colour distribution for 1800 objects with galactocentric radii from 240\arcsec to 390\arcsec and $T_{1o}=$20.0 to 24.0 in $NGC~4486$. Dots indicate the colour peaks detected by the $PFR$ routine. Vertical dashed lines indicate the $(C-T_{1})$ components of the $TVP$. Colours have been corrected according to the values listed in Table 2.
}
    \label{fig:fig17}
\end{figure}

\begin{figure}
	\includegraphics[width=\columnwidth]{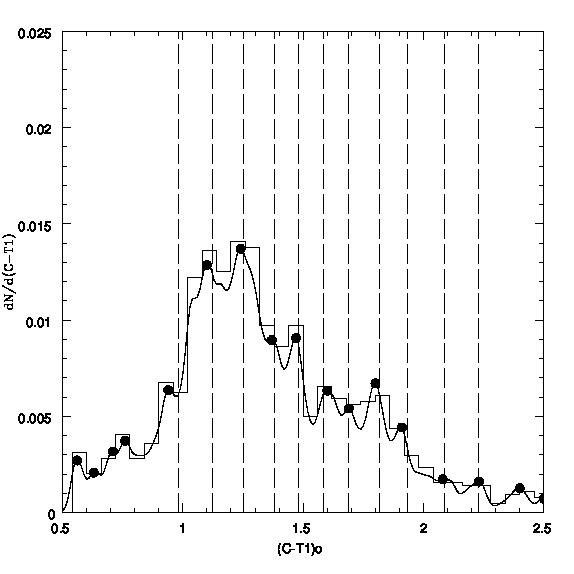}
\vspace{-0.5cm}
    \caption{Globular cluster $(C-T_{1})o$ colour distribution for 1100 objects with galactocentric radii from 240\arcsec to 390\arcsec and $T_{1o}=$22.0 to 23.7, in $NGC~4486$.  Dots indicate the colour peaks detected by the $PFR$ routine. Vertical dashed lines indicate the $(C-T_{1})$ components of the $TVP$. Colours have been corrected according to the values listed in Table 2.
}
    \label{fig:fig18}
\end{figure}

 Finally, the $GCCDs$ displayed in Fig.~\ref{fig:fig17} and Fig.~\ref{fig:fig18}, correspond to $T_{1o}$ magnitude ranges of 20.0 to 24.0 and 
 22.0 to 23.7, respectively. This last diagram, defined within the $T_{1o}$ magnitude range where the pattern is better defined, displays the feature at [1.04] and also, coincidences with all the   $TVP$ components, starting with the peak at [1.12].
%--------------------------------------------------------------------------
%---------------------------------------------------------------------------
%---------------------------------------------------------------------------

\subsection{The GCCD in NGC 4486 for galactocentric radii between 90\arcsec and 390\arcsec.}
\label{ssection5c}
\begin{figure}
	\includegraphics[width=\columnwidth]{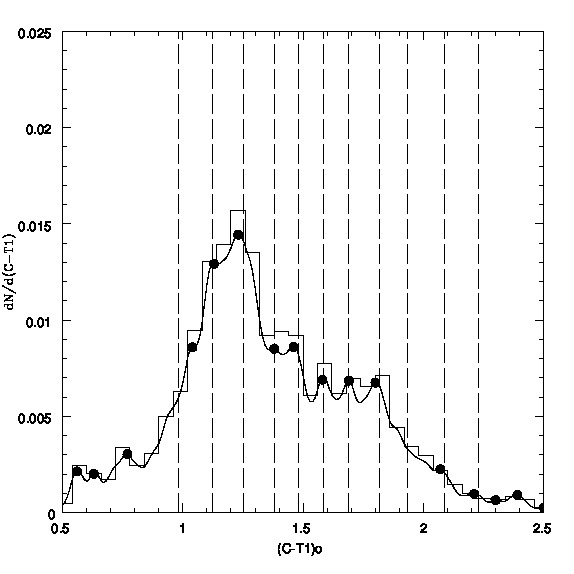}
\vspace{-0.5cm}
    \caption{Globular cluster $(C-T_{1})o$ colour distribution for 3848 objects with galactocentric radii from 90\arcsec to 390\arcsec and $T_{1o}=$20.0 to 24.0 in $NGC~4486$. Dots indicate the colour peaks detected by the $PFR$ routine. Vertical dashed lines indicate the $(C-T_{1})$ components of the $TVP$. Colours have been corrected as explained in the text.
}
    \label{fig:fig19}
\end{figure}

\begin{figure}
	\includegraphics[width=\columnwidth]{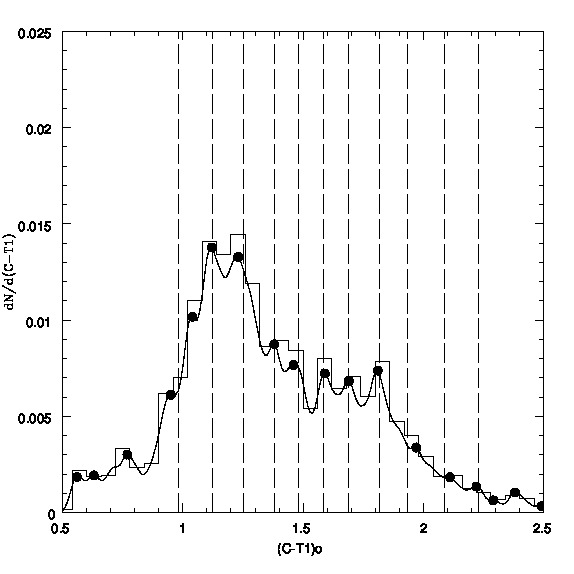}
\vspace{-0.5cm}
    \caption{Globular cluster $(C-T_{1})o$ colour distribution for 2389 objects with galactocentric radii from 90\arcsec to 390\arcsec and $T_{1o}=$22 to 23.7 in $NGC~4486$. Dots indicate the colour peaks detected by the $PFR$ routine. Vertical dashed lines indicate the $(C-T_{1})$ components of the $TVP$. Colours have been corrected as explained in the text.
}
    \label{fig:fig20}
\end{figure}

 The composite $GCCDs$, after combining the results discussed in the two previous subsections, and for two magnitude ranges ($T_{1o}=$20.0 to 24.0 and $T_{1o}=$22 to 23.7), are presented in Fig.~\ref{fig:fig19} 
($rms=$0.019) and Fig.~\ref{fig:fig20} ($rms=$0.017), respectively. Both diagrams display essentially the same colour patterns. This region covers an area about seven times larger than a typical $ACS$ frame
 and includes a contamination level by field interlopers between 10 and 15 percent at a limiting magnitude $T_{1}=$23.7 (see Section 4), i. e., 200 to 250 objects. These field interlopers may produce 
 spurious peaks that eventually smear-out the $TVP$ pattern.
%--------------------------------------------------------------------------
%----------------------------------------------------------------------------------------------------------
%----------------------------------------------------------------------------------------------------------

\begin{figure}
	\includegraphics[width=\columnwidth]{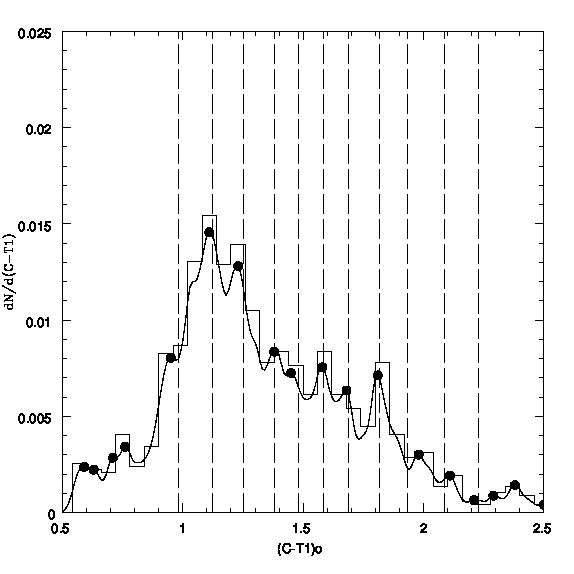}
\vspace{-0.5cm}
    \caption{Globular cluster $(C-T_{1})o$ colour distribution for 1201 objects with galactocentric radii from 90\arcsec to 390\arcsec and $T_{1o}=$22 to 23.7 and position angles from 0\degr to
 180\degr in $NGC~4486$. Dots indicate the colour peaks detected by the $PFR$ routine. Vertical dashed lines indicate the $(C-T_{1})$ components of the $TVP$. Colours have been corrected as explained in the text.
}
    \label{fig:fig21}
\end{figure}

\begin{figure}
	\includegraphics[width=\columnwidth]{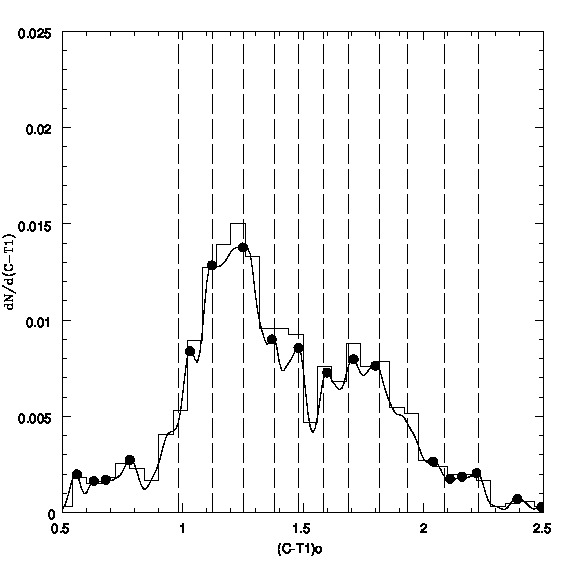}
\vspace{-0.5cm}
    \caption{Globular cluster $(C-T_{1})o$ colour distribution for 1181 objects with galactocentric radii from 90\arcsec to 390\arcsec and $T_{1o}=$22 to 23.7, and position angles from 180\degr to 360\degr in $NGC~4486$. Dots indicate the colour peaks detected by the $PFR$ routine. Vertical dashed lines indicate the $(C-T_{1})$ components of the $TVP$. Colours have been corrected as explained in the text.
}
    \label{fig:fig22}
\end{figure}

The persistence of the $TVP$ as a function of spatial distribution is confirmed in Fig.~\ref{fig:fig21} and Fig.~\ref{fig:fig22}, corresponding to the $GCCDs$ for objects with $T_{1o}=$22.0 to 23.7 and
position angles from 0\degr to 180\degr~ ($rms=$0.018) and 180\degr to 360\degr~ ($rms=$0.017), respectively.
%----------------------------------------------------------------------------------------------------------
%----------------------------------------------------------------------------------------------------------
%----------------------------------------------------------------------------------------------------------

On the other side, the behaviour of the $TVP$ for different ranges of the $T_{1o}$ magnitude, after adopting the colour shifts listed in Table 2, is displayed in Fig.~\ref{fig:a1} to Fig.~\ref{fig:a5}. The first of these diagrams corresponds to 946 objects with $T_{1o}$ from 20.0 to 22.0. This $GCCD$ has a dominant and single blue peak [$(C-T_{1})_{o}=$1.25]. As noted previously, some degree of
 contamination by ultra compact objects can be expected in this range of magnitude. In turn, Fig.~\ref{fig:a2} ($T_{1o}=$ 22.0 to 22.7; 816 objects), Fig.~\ref{fig:a3} ($T_{1o}=$ 22.7 to 23.25; 777 objects), and Fig.~\ref{fig:a4} ($T_{1o}=$ 23.25 to 23.7; 785 objects), show the presence of seven to eight $TVP$ features. These ranges  in $T_{1}$ magnitude were set in order to include an approximately similar number of cluster candidates ($\approx$800 objects). The blue peak at (1.04) is seen in both Fig.~\ref{fig:a2} and Fig.~\ref{fig:a3}.

Fig.~\ref{fig:a5} corresponds to 1212 objects with $T_{1o}$ from 23.7 to 24.5. In this sample the $TVP$ is not detectable as a consequence of the larger photometric errors as well as the presumably
increasing number of field interlopers. 

\begin{figure}
	\includegraphics[width=\columnwidth]{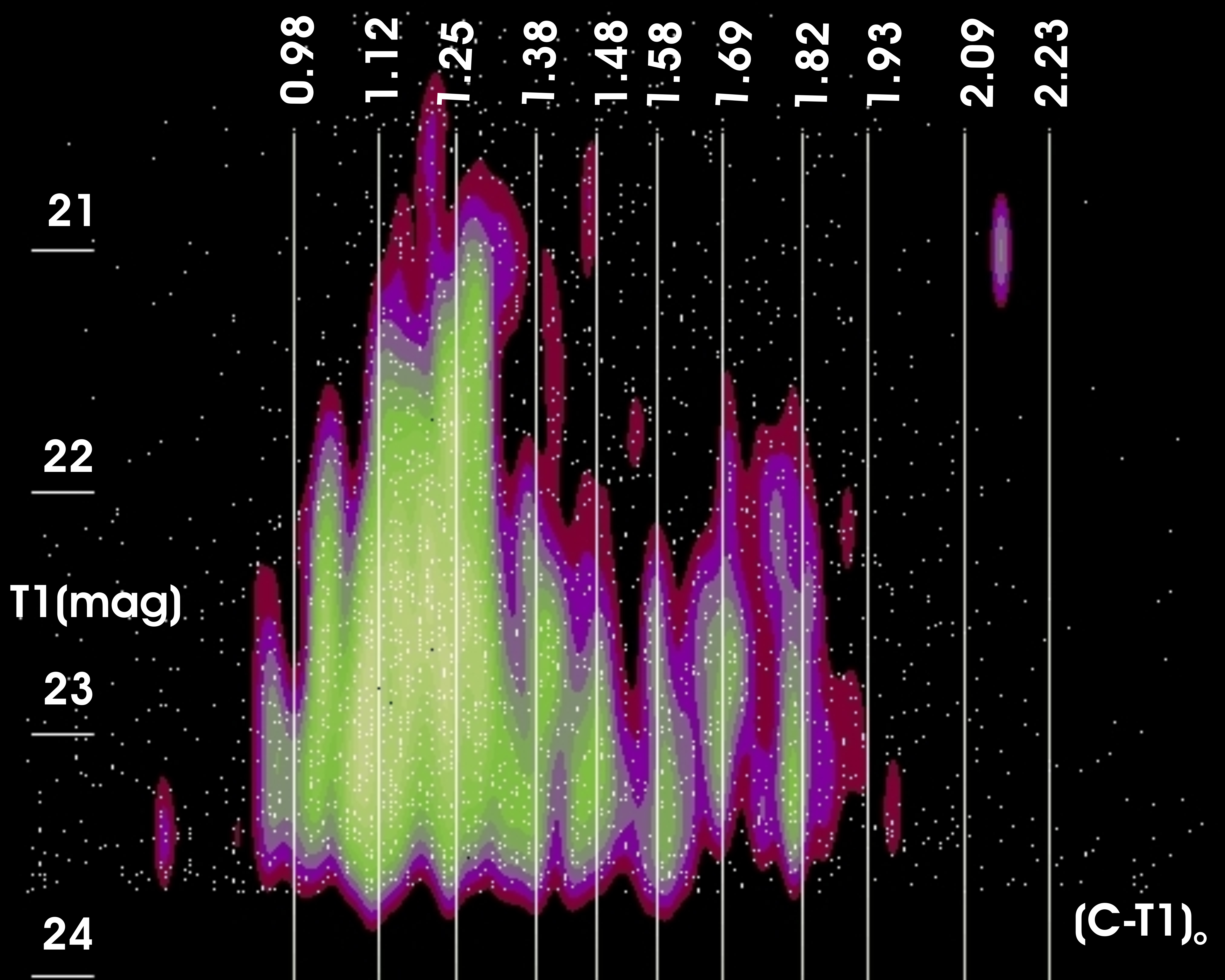}
\vspace{-0.5cm}
    \caption{Smoothed colour-magnitude diagram for 3324 $GCs$ with galactocentric radii from 90\arcsec to 390\arcsec and $T_{1o}=$20.0 to 23.7,
 and adopting the colour shifts listed in Table 2. The vertical lines correspond to the $TVP$ colours. White dots represent individual $GCs$. The contour of the convolving kernel is
 shown at upper right.
}
    \label{fig:fig23}
\end{figure}

In turn, Fig.~\ref{fig:fig23} shows the smoothed colour-magnitude diagram (similar to that introduced in Fig.~\ref{fig:fig6}). In this diagram, that includes 3324 $GCs$, we set a magnitude cut off at $T_{1o}=$23.70, as fainter clusters do not show the $TVP$ (see Fig.~\ref{fig:fig12}). The contour of the assymetric convolving kernel (0.25 mag. in $T_{1o}$ and 0.019 mag. in $(C-T_{1}$)  is displayed at upper right. Again, as in Fig.~\ref{fig:fig6}, the $GCCD$ seems highly structured and in concordance with the $TVP$ colours.

Besides the two dominant blue peaks, at [1.12] and [1.25], Fig.~\ref{fig:fig23} suggests the presence of several bluer components, being the feature at [1.04] the most prominent one.

\section{The integrated GC Luminosity Functions of the TVP components in NGC 4486.}
\label{s6}
 The integrated luminosity function ($LF$ in what follows) corresponding to 2931 $GCs$ with $(C-T_{1})$ colours from 0.93 to 2.28 (and normalized by the total number of clusters) is displayed in Fig. A6. As expected, this distribution can be properly represented with a Gaussian function characterized by a mean magnitude $T_{1o}$=23.3 and a dispersion $\sigma_{T_{1}}=$ 1.3 mag, also shown in this and in figures A7 to A16. The turn-over magnitude is consistent with that derived by \citet{Villegas2010}. All these diagrams include the size of the convolving kernel (adopted for Figs. 6 and 23) which is about 5 percent of the total range in the $T_{1o}$ magnitude.

 In order to explore the nature of the colour peaks that define the $TVP$, we analysed the characteristics of both the distribution of the $GCs$ on the sky and of their integrated $T_{1o}$ luminosity functions.  This analysis was performed adopting 0.1 mag. colour windows centered at the detectable $TVP$ components and also at the [1.04] feature.

The nature of the [1.04] colour peak is intriguing. This feature, and in general objects bluer than $(C-T_{1})=$ 1.17, share similar and very shallow spatial distributions. Their $LFs$, shown in Fig. A7, A8, and A9 are also similar and steeper than the Gaussian that gives a good representation of the whole $GC$ sample. It is tempting to suggest that, at least a fraction, of this population of $GC$ candidates are in fact members of the so called "Virgo Intra Cluster GC population" \citep[e.g.][]{Ko2018} although this cannot be proven just on photometric data.

The remaining $TVP$ components show a variable degree of concentration towards the galaxy centre, that increases with redder colours. 

In general, the $LFs$  shown in Fig. A7 to A16 are  broadly consistent with the Gaussian approximation shown in Fig. A6 (i.e., the observed counts are within $\approx$ 3 times the expected Poissonian deviations). However, the clarification of the nature of some "clumps" in these figures will require further photometric and spectroscopic data.
%--------------------------------------------------------------------------
%---------------------------------------------------------------------------
\section{Conclusions}
\label{sec7}

In summary, this paper reports that:
\begin{enumerate}
\item A revision of the $TVP$ colours, after increasing the $GCs$ sample by a factor of three, compared to
 that discussed in $F2017$, confirms that the pattern is a solid feature detectable in the 27 brightest
 Virgo galaxies, including 12 giant galaxies not considered in that work.

\item  After grouping the $GCs$ in these 27 galaxies in three independent samples, each with approximately the
 same number of objects ($\approx$ 1400 clusters), the $GCCDs$  exhibit between six and eight colour peaks 
 that are compatible with the components of the $TVP$ presented in $F2017$.

\item At least seven of the $TVP$ components are readily detectable in $NGC~4486$, for $GCs$ having $ACS$ photometry, 
and $R_{gal}$ from 40\arcsec to 90\arcsec . 

\item In turn, a new analysis of the  photometry in the $C$ and $T_{1o}$ filters (Washington system) presented in 
 $FFG07$ for $GCs$ in $NGC~4486$, indicates that the $TVP$ is detectable on the $(C-T_{1})_{o}$ colours in a
 galactocentric range $R_{gal}$ from 90$\arcsec$ to 240$\arcsec$ without requiring any colour shift. This $GC$ sample
 includes some 2000 objects and does not overlap with the $ACS$ clusters of the inner regions of the galaxy.

\item Clusters with $R_{gal}$ from 240$\arcsec$ to 390$\arcsec$ require redward colour shifts in order to match the colour patterns in
 different galactocentric annuli in $NGC~4486$ to the $TVP$ colours . As a whole, these shifts are compatible with the $GCs$
 galactocentric colour gradient determined in \citet{Forte2012} and suggest that these clusters have lower chemical abundances
 than those that defined the $TVP$.

\item The adoption of a distance modulus of 31.1 for the Virgo cluster \citep[e.g.][]{Tonry2001}, then indicates that the $TVP$ is
detectable, at least, in a galactocentric range of 3 to 30 Kpc (40\arcsec to 390\arcsec).

\item The presence of the $TVP$ is not only detectable at different galactocentric radii but also when the $GCs$ sample in 
 the whole galactocentric range from 90\arcsec to 390\arcsec is split as a function of position angle (e.g., from 0\degr to 180\degr and 
 from 180\degr to 360\degr).

\item  The detection of the $TVP$ on top of the clusters colour gradient in $NGC~4486$, suggests that these features
 were eventually imprinted during the dissipative collapse stage of the formation of the galaxy.

\item Globally, the $TVP$ components are detectable on the two cluster (blue and red) sub populations that define
 the broad $GCs$ colour "bi-modality" in $NGC~4486$ (see, for example, $FFG07$).

\item The presence of the $TVP$ pattern on both the $(g-z)o$ and $(C-T_{1})_{o}$ colours, and in different $GCs$ samples,
gives a strong argument to reject instrumental effects as the origin of that pattern. Another argument in the 
same sense is provided by the fact that the $TVP$ is not evident neither for $GCs$ belonging to galaxies fainter than $M_{g}=$-18.2,
 nor for clusters inside a galactocentric radius of 40\arcsec in giant galaxies. Other causes, as contamination by field interlopers
 or statistical fluctuations were already rejected in $F2017$.

\item The existence of two blue $GCs$ components, with $(g-z)o$=0.85 and 0.95 respectively, and most evident for clusters
in moderately bright galaxies (i.e. $M_{g}=$-20.2 to -19.2), is consistent with a similar situation detectable in the outer
 regions of $NGC~4486$ (with corresponding colour peaks at $(C-T_{1})o=$ 1.12 and 1.25).

\item Regarding the eventual "universality" of the $TVP$ (and as noted in $F2017$), each $GC$ system might have clusters that
 are part of its own history (including galaxy mergers, etc.) as well as other clusters that eventually formed in connection
 with the $TVP$. Because of this, the $TVP$ detectability would depend on each particular case. However, the pattern emerges even
 in massive galaxies with presumably complex histories, as $NGC~4486$, discussed in this paper.
\end{enumerate}
    A tentative explanation for the colour patterns, given in $F2017$, suggests that an external mechanism has been able to
 modulate the $GCs$ formation over supra-galactic spatial scales in a kind of $\bf{viral~process}$, at high redshifts, and following
 what seems to be a temporal sequence.  If this is the situation, the effect of such a putative phenomenon might also
 have left its signature early on the diffuse stellar population of these galaxies.

The lack of substantial evidence supporting a given candidate mechanism ($AGN$ activity ? sub-cluster merger events ?) that could
 be responsible for the $GCs$ colour pattern, defines the current situation as a $\bf{supra-galactic~conundrum}$ ("...a question or
 problem having only a conjectural answer"; Merriam-Webster dictionary) that deserves further exploration.

\section*{Acknowledgements}
This work was funded with grants from Consejo Nacional de Investigaciones Cient\'ificas y T\'ecnicas de la Rep\'ublica Argentina $(CONICET)$. J.C.F. thanks 
 Dr. Gustavo Corach for his hospitality at the IAM-CONICET. D.G. gratefully acknowledges support from the Chilean Centro de Excelencia en Astrof\'isica y Tecnolog\'ias Afines (CATA) BASAL grant AFB-170002,
and also the financial support from the Direcci\'on de Investigaci\'on y Desarrollo de
la Universidad de La Serena through the Programa de Incentivo a la Investigaci\'on de
Acad\'emicos (PIA-DIDULS).\\
  
\bibliographystyle{mnras}
\bibliography{biblio_Forte.bib}
% if your bibtex file is called example.bib
%%%%%%%%%%%%%%%%%%%%%%%%%%%%%%%%%%%%%%%%%%%%%%%%%%
% Don't change these lines
\bsp	% typesetting comment
\appendix
\section{Additional diagrams.}
\newpage
%---------------------------------------------------------------------------------------------
\begin{figure}
	\includegraphics[width=\columnwidth]{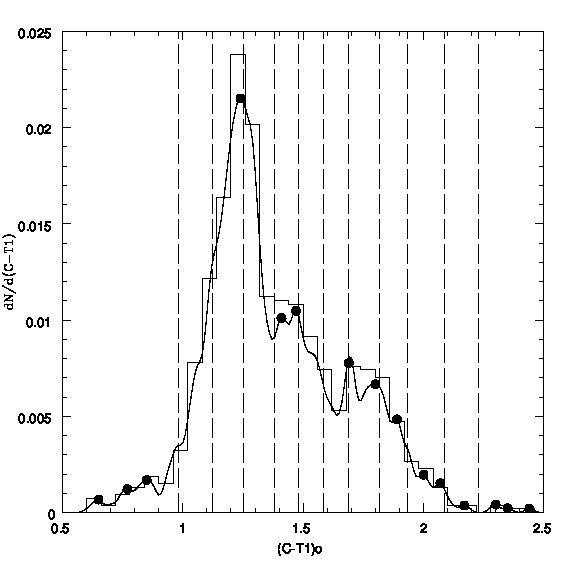}
    \caption{Globular cluster $(C-T_{1})o$ colour distribution for 946 objects with galactocentric radii from 90\arcsec to 390\arcsec and $T_{1o}=$20.0 to 22.0 in $NGC~4486$. Dots indicate the colour peaks detected by the $PFR$ routine. Vertical dashed lines indicate the $(C-T_{1})$ components of the $TVP$. Colours have been corrected as explained in the text.
}
    \label{fig:a1}
\end{figure}
%---------------------------------------------------------------------------------------------
\begin{figure}
	\includegraphics[width=\columnwidth]{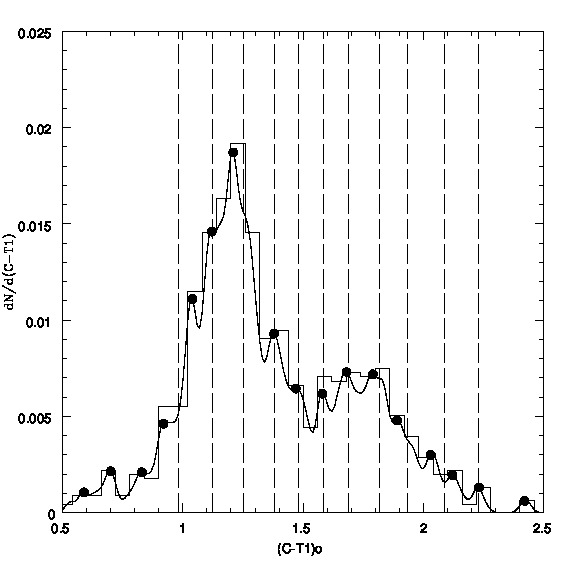} 
    \caption{Similar to Fig. A1 for 816 objects with $T_{1o}=$22.00 to 22.70.
}
    \label{fig:a2}
\end{figure}
%---------------------------------------------------------------------------------------------
\begin{figure}
	\includegraphics[width=\columnwidth]{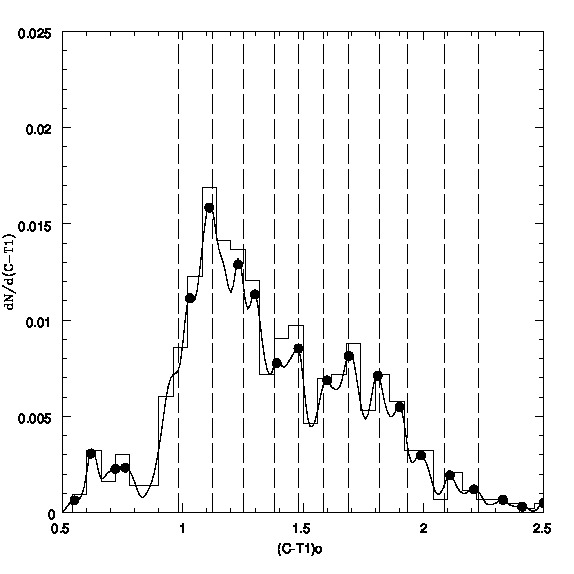}
    \caption{Similar to Fig. A1 for 777 objects with $T_{1o}=$22.70 to 23.25.
}
    \label{fig:a3}
\end{figure}
%---------------------------------------------------------------------------------------------
\begin{figure}
	\includegraphics[width=\columnwidth]{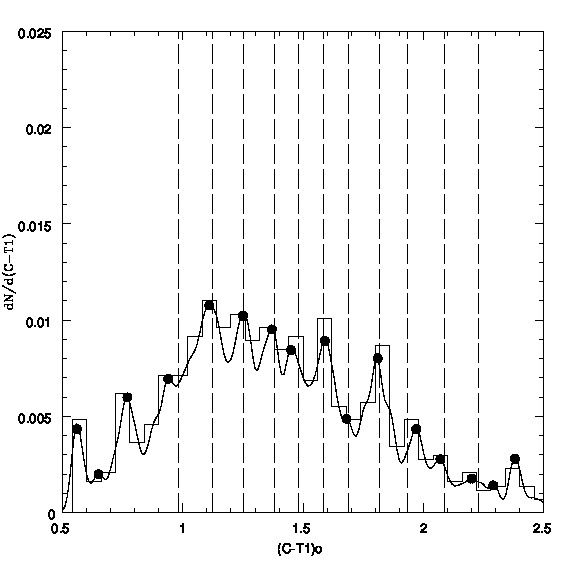}
    \caption{Similar to Fig. A1 for 785 objects with $T_{1o}=$23.25 to 23.70.
}
    \label{fig:a4}
\end{figure}
%---------------------------------------------------------------------------------------------
\begin{figure}
	\includegraphics[width=\columnwidth]{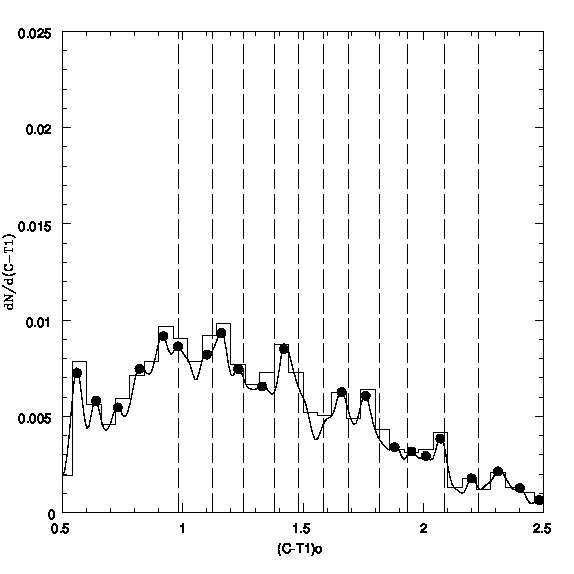}
    \caption{Similar to Fig. A1 for 1212 objects with $T_{1o}=$23.70 to 24.50.
}
    \label{fig:a5}
\end{figure}
%---------------------------------------------------------------------------------------------
\begin{figure}
	\includegraphics[width=\columnwidth]{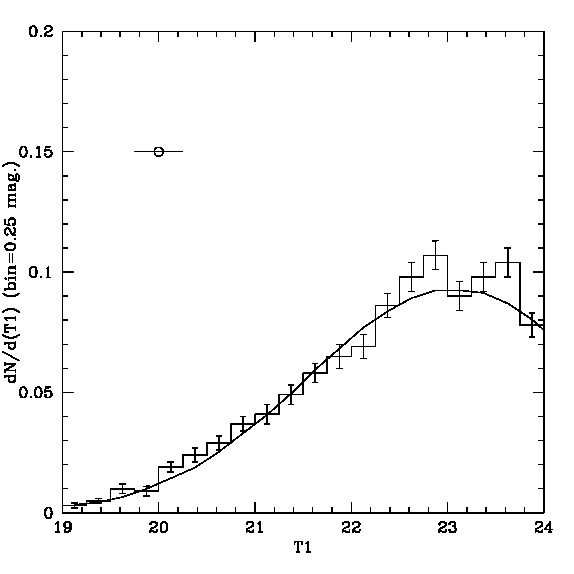}
    \caption{Integrated GCs luminosity function for 2782 GC candidates with galactocentric radii from 90$\arcsec$ to 390$\arcsec$
 and $(C-T_{1})o$ colours from 0.93 to 2.28. The solid line corresponds to a reference Gaussian fit (see text). Vertical bars indicate
 the counting uncertainties (see text).The horizontal bar at upper left shows the size of the convolving kernel used in Figs. 6 and 23.
}
    \label{fig:a6}
\end{figure}
%-------------------------------------------------------------------------------------------------
\begin{figure}
	\includegraphics[width=\columnwidth]{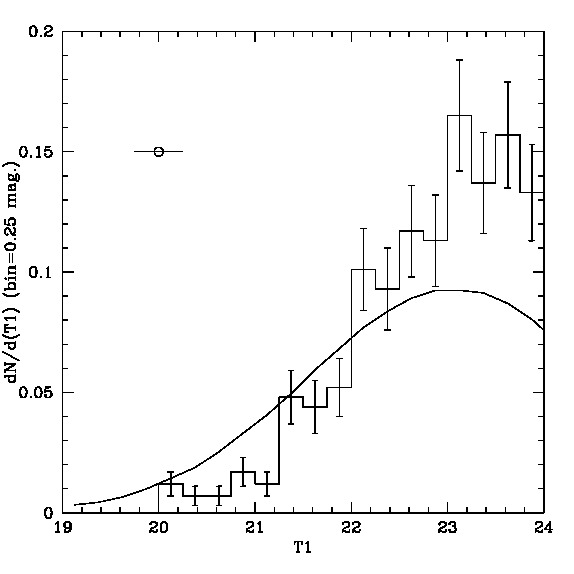}
    \caption{Similar to Fig. A6 but for 275 GC candidates with $(C-T_{1})o$ colours from 0.93 to 1.03.
}
    \label{fig:a7}
\end{figure}
%------------------------------------------------------------------------------------------------------------
\begin{figure}
	\includegraphics[width=\columnwidth]{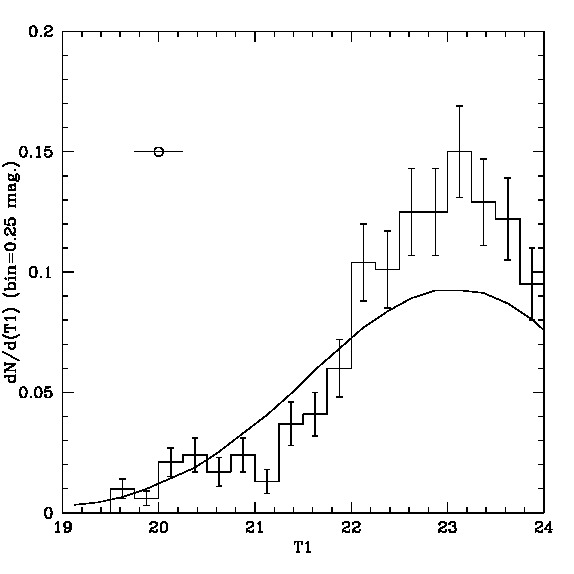}
    \caption{Similar to Fig. A6 but for 352 GC candidates with $(C-T_{1})o$ colours from 0.99 to 1.09.
}
    \label{fig:a8}
\end{figure}
%--------------------------------------------

\begin{figure}
	\includegraphics[width=\columnwidth]{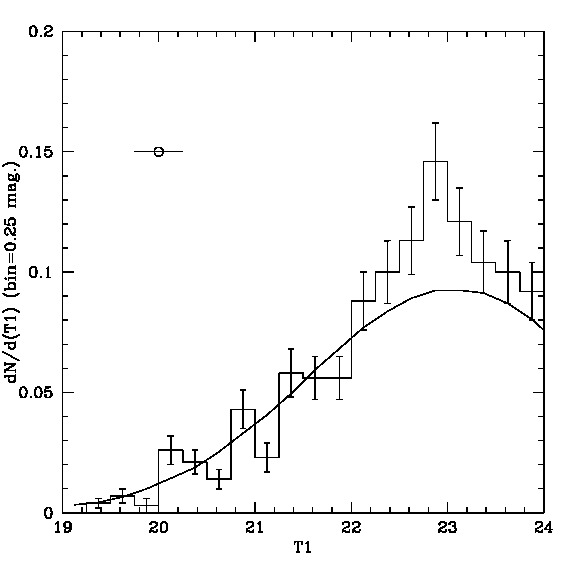}
    \caption{Similar to Fig. A6 but for 516 GC candidates with $(C-T_{1})o$ colours from 1.07 to 1.17.
}
    \label{fig:a9}
\end{figure}
%----------------------------------------------------------------------------------------------------------
\begin{figure}
	\includegraphics[width=\columnwidth]{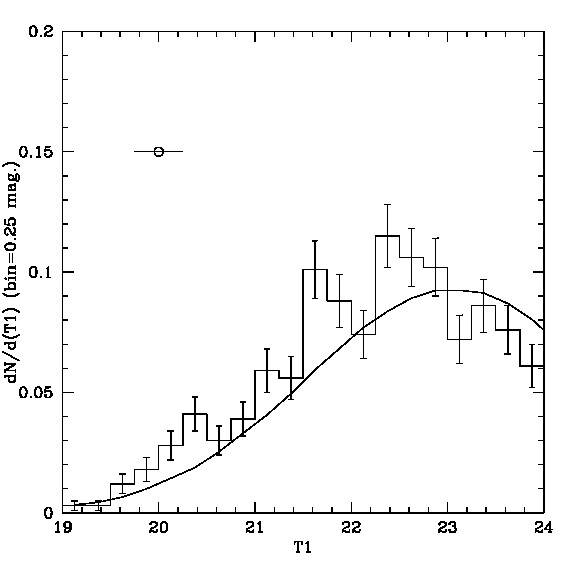}
    \caption{Similar to Fig. A6 but for 599 GC candidates with $(C-T_{1})o$ colours from 1.20 to 1.30.
}
    \label{fig:a10}
\end{figure}
%--------------------------------------------------------------------------------------------------------------
\begin{figure}
	\includegraphics[width=\columnwidth]{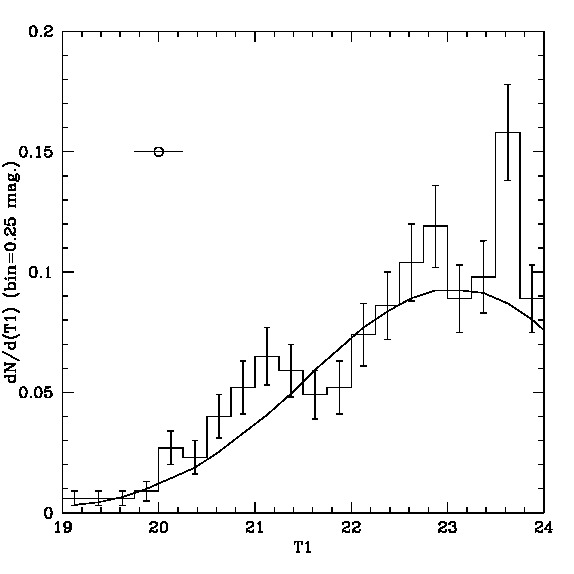}
    \caption{Similar to Fig. A6 but for 361 GC candidates with $(C-T_{1})o$ colours from 1.33 to 1.43.
}
    \label{fig:a11}
\end{figure}
%---------------------------------------------------------------------------------------------------------------
\begin{figure}
	\includegraphics[width=\columnwidth]{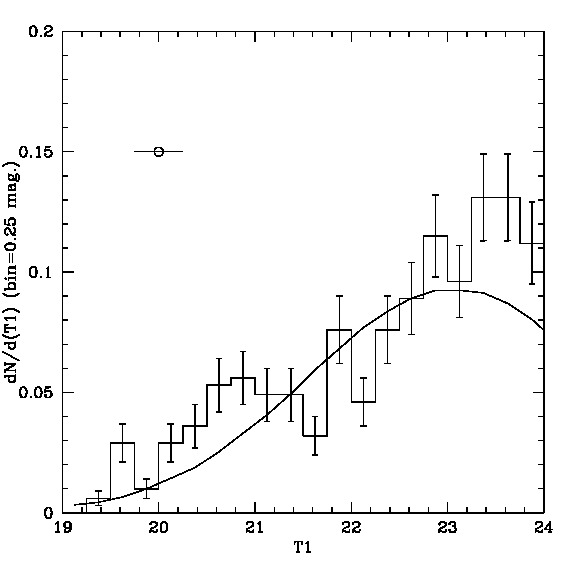}
    \caption{Similar to Fig. A6 but for 337 GC candidates with $(C-T_{1})o$ colours from 1.43 to 1.53.
}
    \label{fig:a12}
\end{figure}
%\newpage
%---------------------------------------------------------------------------------------
\begin{figure}
	\includegraphics[width=\columnwidth]{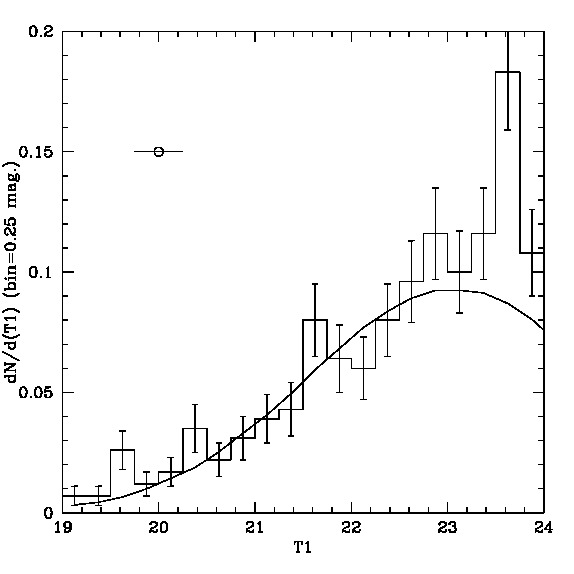}
    \caption{Similar to Fig. A6 but for 277 GC candidates with $(C-T_{1})o$ colours from 1.53 to 1.63.
}
    \label{fig:a13}
\end{figure}
%----------------------------------------------------------------------------------------
\begin{figure}
	\includegraphics[width=\columnwidth]{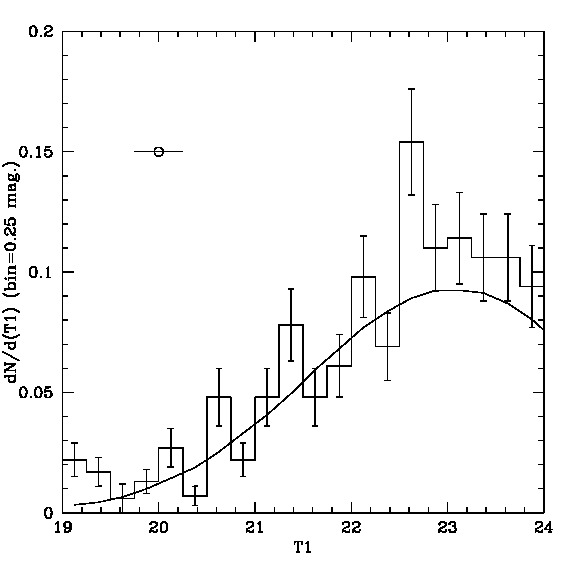}
    \caption{Similar to Fig. A6 but for 272 GC candidates with $(C-T_{1})o$ colours from 1.64 to 1.74.
}
    \label{fig:a14}
\end{figure}
%--------------------------------------------
\begin{figure}
	\includegraphics[width=\columnwidth]{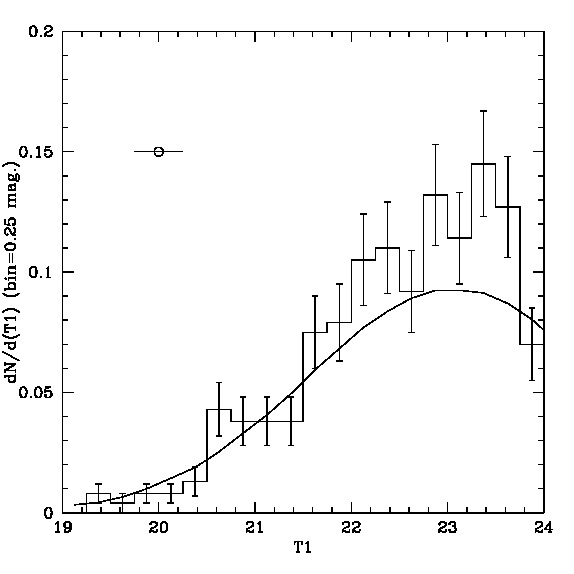}
    \caption{Similar to Fig. A6 but for 253 GC candidates with $(C-T_{1})o$ colours from 1.77 to 1.87.
}
    \label{fig:a15}
\end{figure}
%--------------------------------------------
\begin{figure}
	\includegraphics[width=\columnwidth]{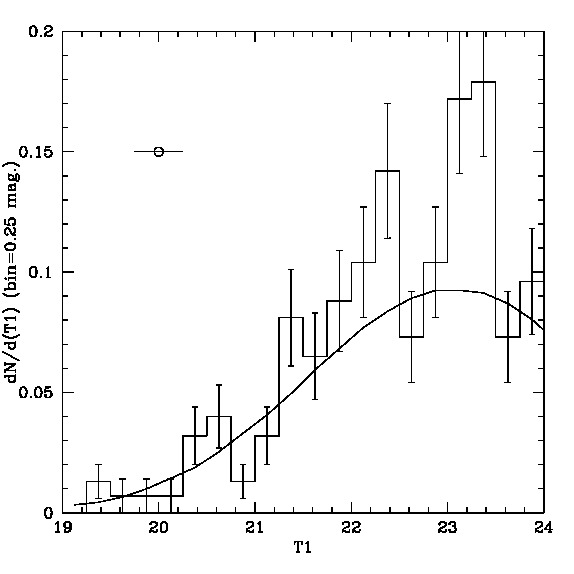}
    \caption{Similar to Fig. A6 but for 149 GC candidates with $(C-T_{1})o$ colours from 1.88 to 1.98.
}
    \label{fig:a16}
\end{figure}
%--------------------------------------------
\label{lastpage}
\end{document}